\documentclass[fleqn,10pt]{wlscirep}
\usepackage[utf8]{inputenc}
\usepackage[T1]{fontenc}
\usepackage{textcomp}
\usepackage{hyperref}
\usepackage{pdfpages}

\title{Learning the rules of peptide self-assembly through data mining with large language models}
\author[1,2]{Zhenze Yang}
\author[3]{Sarah K. Yorke}
\author[3*]{Tuomas P. J. Knowles}
\author[1,4,5,*]{Markus J. Buehler}
\affil[1]{Laboratory for Atomistic and Molecular Mechanics, Department of Civil and Environmental Engineering, Massachusetts Institute of Technology, 77 Massachusetts Ave. Room 1-165, Cambridge, 02139, MA, USA}
\affil[2]{Department of Materials Science and Engineering, Massachusetts Institute of Technology, 77 Massachusetts Ave., Cambridge, 02139, MA, USA}
\affil[3]{Yusuf Hamied Department of Chemistry, University of Cambridge, Lensfield Road, Cambridge, CB2 1EW, UK}
\affil[4]{Center for Computational Engineering, Massachusetts Institute of Technology, 77 Massachusetts Ave., Cambridge, 02139, MA, USA}
\affil[5]{Center for Materials Science and Engineering, Massachusetts Institute of Technology, 77 Massachusetts Ave., Cambridge, 02139, MA, USA}
\affil[*]{mbuehler@mit.edu, tpjk2@cam.ac.uk}

\begin{abstract}
\noindent Peptides are ubiquitous and important biologically derived molecules, that have been found to self-assemble to form a wide array of structures. Extensive research has explored the impacts of both internal chemical composition and external environmental stimuli on the self-assembly behaviour of these systems. However, there is yet to be a systematic study that gathers this rich literature data and collectively examines these experimental factors to provide a global picture of the fundamental rules that govern protein self-assembly behavior. In this work, we curate a peptide assembly database through a combination of manual processing by human experts and literature mining facilitated by a large language model. As a result, we collect more than 1,000 experimental data entries with information about peptide sequence, experimental conditions and corresponding self-assembly phases. Utilizing the collected data, ML models are trained and evaluated, demonstrating excellent accuracy (>80\%) and efficiency in peptide assembly phase classification. Moreover, we fine-tune our GPT model for peptide literature mining with the developed dataset, which exhibits markedly superior performance in extracting information from academic publications relative to the pre-trained model. We find that this workflow can substantially improve efficiency when exploring potential self-assembling peptide candidates, through guiding experimental work, while also deepening our understanding of the mechanisms governing peptide self-assembly. In doing so, novel structures can be accessed for a range of applications including sensing, catalysis and biomaterials.
\end{abstract}
\begin{document}

\flushbottom
\maketitle
%
%
\thispagestyle{empty}

\section*{Introduction}
Self-assembly is a ubiquitous phenomenon in nature that plays a critical role in the formation of hierarchical biomaterials. A wide array of building blocks ranging from organic molecules and proteins to nucleic acids and amphiphilic compounds can self-assemble into nano- and micro- scale structures, driven by various molecular interactions \cite{STEPHANOPOULOS2013912, Knowles2011, doi:10.1126/science.1074200}. Among these building blocks, peptides present unique structural and functional characteristics with their simple yet versatile chemical compositions and non-covalent, weak intermolecular interactions. In the past few decades, a rich diversity of self-assembled peptide-based nanostructures have been reported, including tubes, fibers, ribbons, plates and spheres \cite{doi:10.1126/science.1205962, Zhang2003, Hartgerink1996, Levin2014, doi:10.1021/ja052644i, doi:10.1021/ja9937831}. These assembled peptide serve as structured, functional biomaterials, spanning a broad spectrum of applications from drug delivery systems \cite{Sis2019} and tissue engineering\cite{https://doi.org/10.1002/jbm.a.35638}, to catalysis \cite{HAN2021101295} and electronics \cite{doi:10.1021/acs.bioconjchem.5b00497}. \\
\\
The self-assembly behaviour of peptides is mediated by the underlying thermodynamics and kinetics of the systems \cite{C6CS00176A}. Peptides can adopt specific organisations, such as supra-molecular $\alpha$-helices, $\beta$-sheets, and $\beta$-hairpins, which are key building blocks of specific assembled nanostructures \cite{LI2022268}. As a result, extensive studies have demonstrated that one can use a range of factors to manipulate a system into forming specific secondary structures. These include intrinsic parameters, such as peptide sequence \cite{doi:10.1021/jz2010573} and chemical modifications \cite{Levin2014}, as well as extrinsic factors such as pH \cite{doi:10.1021/acsabm.2c00188}, temperature \cite{doi:10.1021/ja0764862}, solvent type \cite{doi:10.1021/jacs.0c03425, doi:10.1021/la800942n, doi:10.1021/nn404237f}, and concentration \cite{doi:10.1021/ja1025535}. However, a systematic study collecting this literature data and exploring how these experimental conditions collectively influence the phase diagram of peptide materials is yet to be conducted. Something of this nature would be a groundbreaking tool, helping to streamline experimental investigations through identifying key mediators of self-assembly. \\
\\
Machine learning (ML) and artificial intelligence (AI) have emerged as unique and invaluable tools across various fields beyond computer science. For example, designing new proteins with target functionalities using ML and AI techniques has revolutionized bioengineering \cite{Watson2023, NI20231828, YANG2023105098, https://doi.org/10.1002/adfm.202311324, D4DD00013G}. A relevant field is polypeptide self-assembly, in which both classical ML algorithms (random forest \cite{Batra2022, Xu2023}) and deep learning (DL) approaches (graph neural networks \cite{doi:10.1021/acsbiomaterials.3c01001} or transformer-based models \cite{https://doi.org/10.1002/advs.202301544}) have been applied to predict the aggregation propensity of polypeptide materials given different amino acid sequences. It is challenging to generate large experimental datasets with synthetic molecules, thus the datasets used for training these models are generally curated from coarse-grained molecular dynamics (MD) simulations \cite{Batra2022, Xu2023, https://doi.org/10.1002/advs.202301544} or statistical algorithms based on physicochemical properties \cite{Fernandez-Escamilla2004, 10.1093/nar/gkv359, https://doi.org/10.1002/prot.25594}. Compared to experimental approaches, computational methods offer high efficiency but are constrained by their accuracy and the range of intrinsic/extrinsic parameters to which they are applicable. In addition, the aggregation propensity, a commonly predicted metric in protein science, is defined based on surface area and reflects only the capacity of proteins to aggregate, not the structural diversity of these aggregates. Therefore, collecting comprehensive experimental data is crucial for understanding the self-assembly behaviors of peptide materials and laying the groundwork for data-driven research. \\
\\
When it comes to data collection from the literature, often referred to as literature mining, it becomes an intractable task for human experts to review all relevant papers due to the vast volume of publications. The recent advent of language models such as large language models (LLMs) \cite{zhao2023survey}, introduces a novel approach to mining academic texts written by researchers and collecting target information in a fast, automatic and systematic manner. These  LLMs, built based on graph-forming transformer architectures \cite{10.5555/3295222.3295349}, possess the capability to capture ultra-long distance relationships within text documents. With massive training across all types of text resources, LLMs attain the ability to comprehend the general context of human languages, making them versatile for a wide range of downstream language tasks such as graph reasoning \cite{buehler2024acceleratingscientificdiscoverygenerative} and text mining \cite{brown2020language, touvron2023llama}. Empowered by LLMs, numerous materials research studies have succeeded in extracting information about materials composition and properties from the abstract texts of academic publications \cite{Tshitoyan2019, Gupta2022, foppiano2024mining}. However, due to their concise, well-structured and clear composition, abstracts can be easily processed by language models, unlike the more complex main texts with sparse and mixed-modality information (text, images, captions, references, etc.). Therefore, only a limited number of studies have undertaken literature mining of the primary texts of massive publications~\cite{Luu2023BioinspiredLLM:Materials,buehler2024cephalomultimodalvisionlanguagemodels,buehler2024preflexorpreferencebasedrecursivelanguage}, and such work has utilized more complex processing steps to extract knowledge from papers and to then use for training LLMs. Abstracts provide only a minimal amount of information, and typically lack detailed experimental procedures and results, both of which are essential for understanding and replicating self-assembly processes, thereby limiting their utility in mining for comprehensive literature data. \\
\\
In this work, we curate a dataset for self-assembling short peptides with information about assembled phase and corresponding experimental conditions. This dataset is derived from publications selected from a previously established peptide self-assembly database, SAPdb \cite{MATHUR2021104391}, with the information being extracted by human experts. With more than 1,000 data entries from the database, we are able to train and compare ML algorithms that can make accurate phase predictions based on the peptide sequence and experimental conditions. To further automate the process and increase the efficiency of literature mining, we utilize the manually curated dataset to fine-tune the pretrained ChatGPT model (GPT-3.5 Turbo model, referred as GPT model) via OpenAI API which exhibits superior performance in information extraction compared to the original model, without fine-tuning. The model can be used to facilitate data collection for peptide materials, potentially providing access to new self-assembled structures, while also enhancing our comprehension of the underlying principles governing their self-assembly processes.

\section*{Results}
\subsection*{Overall workflow}
\begin{figure}[ht]
\centering
\includegraphics[width=\linewidth]{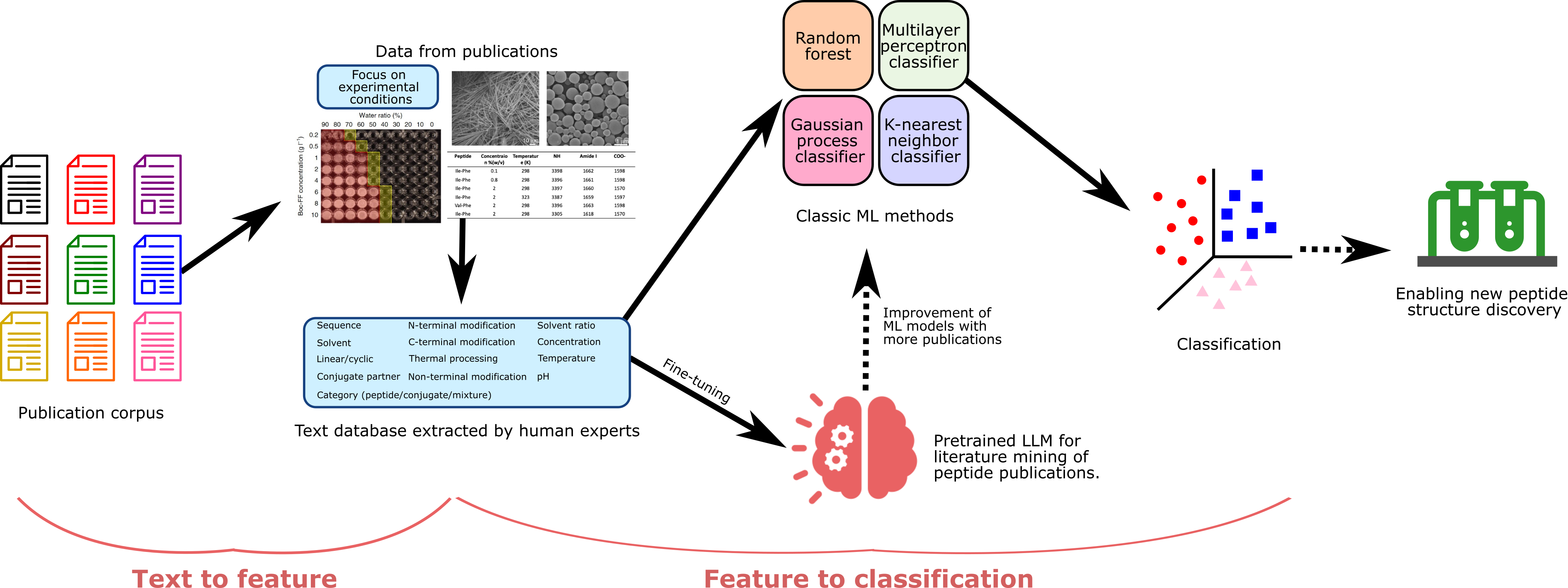}
\caption{\textbf{Overview of the workflow reported in this work}. We first collect PDF files of publications from different journal presses and scientific databases based on the previous polypeptide database SAPdb \cite{MATHUR2021104391}. Here, we extract not only the peptide sequence but also experimental conditions from those previous publications and learn their impacts on the self-assembly phase of polypeptides. The selected publications are read and processed by human experts to curate the database, which is further utilized to train ML algorithms for predicting self-assembled structure from peptide sequences and experimental conditions. We also use the manually curated database to fine-tune a LLM to be specialized in literature mining of polypeptide publications and compare the performance with the same LLM without fine-tuning. The model can be used to extract information for new publications, significantly reducing the time required compared to manual methods employed by human experts. Moreover, by incorporating this new data, we can augment our existing database, thereby further refining and enhancing our ML model for phase prediction.}
\label{fig:workflow}

\end{figure}

The overall workflow of this study is depicted in Fig. \ref{fig:workflow}. We first collect scientific publications related to polypeptide self-assembly from publishers directly, and from scientific databases such as the PubMed database, based on a previous polypeptide database known as SAPdb \cite{MATHUR2021104391}. SAPdb is a collection of 1049 entries of experimentally validated short peptides (di-, tri-peptides) from 301 papers. We screen the entire database based on whether each publication can be adopted into our feature template for ML predictions. In our feature template, we have 9 categorical features and 4 numerical features as shown in Fig. \ref{fig:data_stats}a, b. Academic publications that extend experimental control beyond these specified features are not included in the database, due to their scarcity and limited utility. Following the screening process, we identified a total of 75 publications. All 75 publications are cited in Supporting Information. With these selected publications, we focus on extracting experimental details and the assembled phases from the main texts, which results in a total of 1012 data entries. \\
\\
With the curated database, we then train ML models to predict the self-assembly phases from peptide sequence and experimental parameters.  To obtain an optimal performance in the classification, we compare multiple classic ML algorithms with hyperparameter optimization which are discussed more in detail in the following content (see Section \hyperref[sec:ml_phase_prediction]{ML algorithms for phase prediction}). In contrast to the costly and time-consuming processes of peptide synthesis and characterization, the ML model offers the ability to rapidly classify the peptides with moderate accuracy. However, the manual process of paper reading and data collection is human resource-extensive and time-consuming. Therefore, we further leverage pretrained LLM (GPT-3.5 turbo) to accelerate the process of literature mining. The LLM is engineered to perform a task known as "named entity extraction", enabling the model to efficiently extract key information from text documents based on given targeted entity. To adapt the GPT model to be specialized in understanding scientific writing related to peptide self-assembly, the manually curated dataset is split into training and testing sets, with the training set being used to fine-tune the GPT-3.5 turbo model and testing set being implemented for performance evaluation. We demonstrate that fine-tuning can significantly enhance the performance of information extraction and requires only a small amount of data for transfer learning. \\
\\
With the fine-tuned LLM assistant, we are able to perform efficient literature mining for future publications or research works that are not among selected publications such as those studying longer peptides. This new data, obtained by LLMs, can be further added into our database and used for boosting the performance of our ML model in phase classification. As a result, the proposed approach paves the way for an autonomous workflow capable of continuously collecting data from papers, augmenting the existing dataset, and refining the classification model. Furthermore, this workflow offers the potential to facilitate the design of experiments and screening of promising peptide candidates. 

\subsection*{Features and statistics of dataset}
\begin{figure}[ht]
\centering
\includegraphics[width=\linewidth]{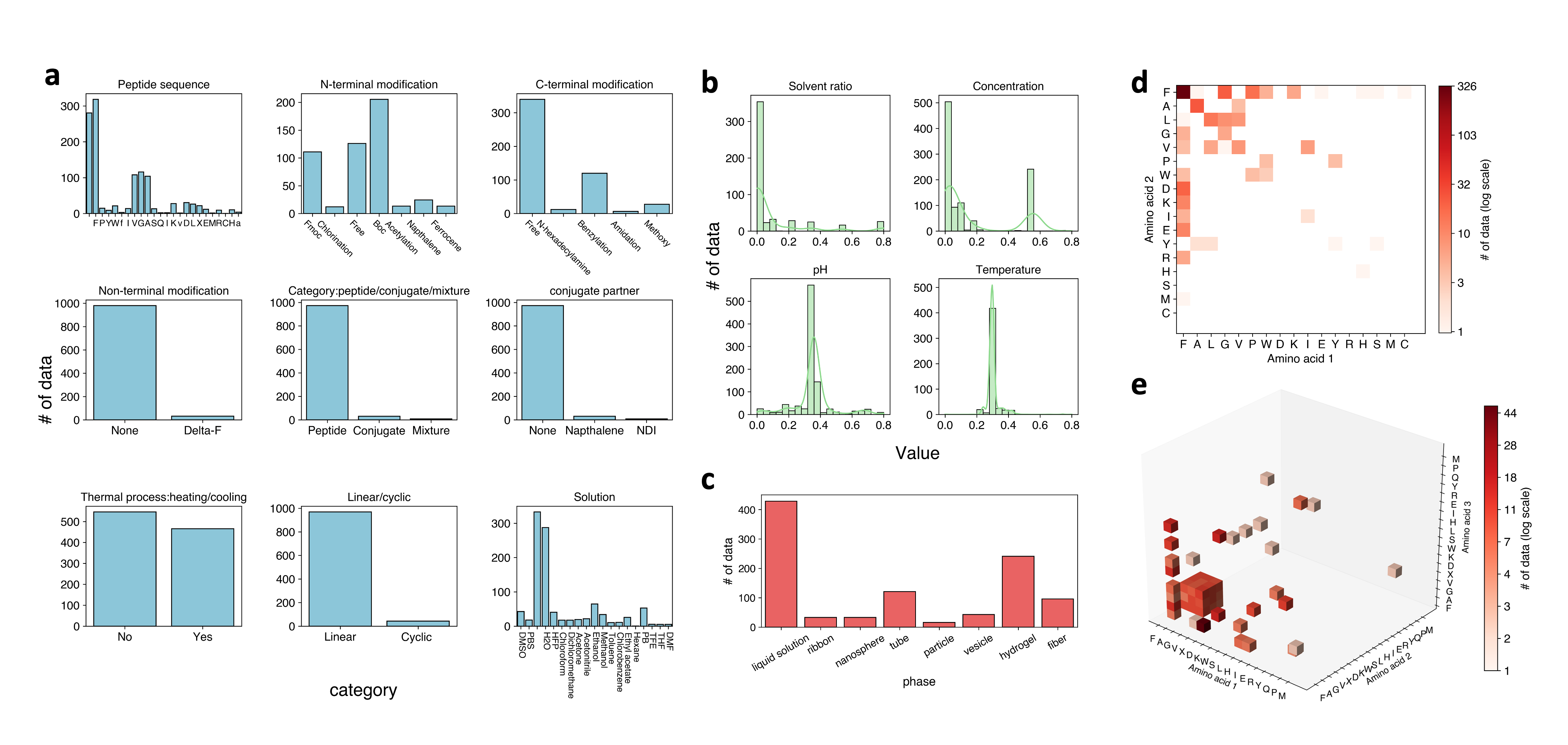}
\caption{\textbf{Dataset statistics}. \textbf{(a)} Histogram of 9 categorical features including "peptide sequence", "N-terminal modification", "C-terminal modification" , "Non-terminal modification", "category: peptide/conjugate/mixture", "conjugate partner", "thermal process: heating/cooling"), "linear/cyclic" and "solution" (the solution environment of the peptide). \textbf{(b)} Histogram of 4 numerical features including "solvent ratio" (the ratio of solvent in the solution), "concentration" (concentration of peptides), "pH" (pH of solution environment), "temperature" (ambient temperature of experiments). All values are normalized between 0 and 1 based on min and max values. \textbf{(c)} Histogram of assembled phases. \textbf{(d)} Occurrence of dipeptide data from academic literature. \textbf{(e)} Occurrence of tripeptide data from academic literature. }
\label{fig:data_stats}
\end{figure}

The statistics of curated polypeptide dataset is shown in Fig. \ref{fig:data_stats}. There are 9 categorical features including: (1) "Peptide Sequence": which is the amino acid sequence of peptides and is specified for dipeptides (length of 2) or tripeptides (length of 3); (2) "N-terminal Modification": which describes chemical modifications, such as the protecting groups, at the N-terminal of peptides; (3) "C-terminal Modification": which involves chemical modifications at the C-terminal of peptides; (4) "Non-terminal Modification": referring to chemical modifications of the R-group; (5) "Category: Peptide/Conjugate/Mixture": indicating whether the peptide system consists of a single peptide, a conjugate, or a mixture; (6) "Conjugate Partner": identifying the conjugated peptide within a conjugate system; (7) "Thermal Process: Heating/Cooling": indicating the presence of temperature changes, such as heating or cooling, during the experimental process; (8) "Linear/Cyclic": distinguishing between linear and cyclic self-assembling peptides; (9) "Solution": detailing the solution environment including information about the solvent and solute. The potential values for each categorical feature, along with their occurrence frequencies in our dataset, are visualized in Fig. \ref{fig:data_stats}a. \\
\\
In addition to categorical features, numerical features also play a crucial role in determining the self-assembly phase of peptides. We here investigate 4 different numerical features as follows: (1) "Solvent ratio;" which describes the volume ratio of solvent in the solution; (2) "Concentration": defined as the concentration (mg/ml) of dissolved peptides; (3) "pH": which is the pH value of the solution environment; (4) "Temperature:" denoting the ambient temperature at which the self-assembly experiment is conducted. The histograms of these numerical features are displayed in Fig. \ref{fig:data_stats}b. All numerical features are normalized between 0 and 1 based on the maximum and minimum values found in the database. \\
\\
The combination of categorical and numerical features constitutes the complete input experimental conditions, encompassing a total of 13 features. The corresponding distribution of output self-assembly phases are shown in Fig. \ref{fig:data_stats}c. As shown in the figure, the most frequent phases seen in the publications is "no-assembly" case which is not included in the original SAPdb database. Nonetheless, these instances are vital for training a ML model for phase prediction, as they contribute valuable non-positive examples to the database. Example data entries in the database are shown in Table S1. \\
\\
We also visualize the occurrence frequency for both dipeptides and tripeptides in our database, as depicted in Fig. \ref{fig:data_stats}d and e. For dipeptides, diphenylalanine (FF) is, as expected, the most studied peptide. FF is the core recognition motif of the Alzheimer's $\beta$-amyloid peptide and can form a wide array phases including hydrogels and hollow nanotubes under different external stimuli\cite{Tamamis2009}. For tripeptides, sequences containing FF are also commonly observed (evident in data on the x-y or y-z planes), with XFF having the highest frequency of occurrence (where X is any given amino acid). This is attributed to the wide range of experimental conditions under which XFF has been examined. Apart from phenylalanine, glycine is the second most investigated amino acids in both dipeptide and tripeptide cases. In addition, research on dipeptides predominates over tripeptides in terms of overall study volume.

\begin{figure}[ht]
\centering
\includegraphics[width=0.8\linewidth]{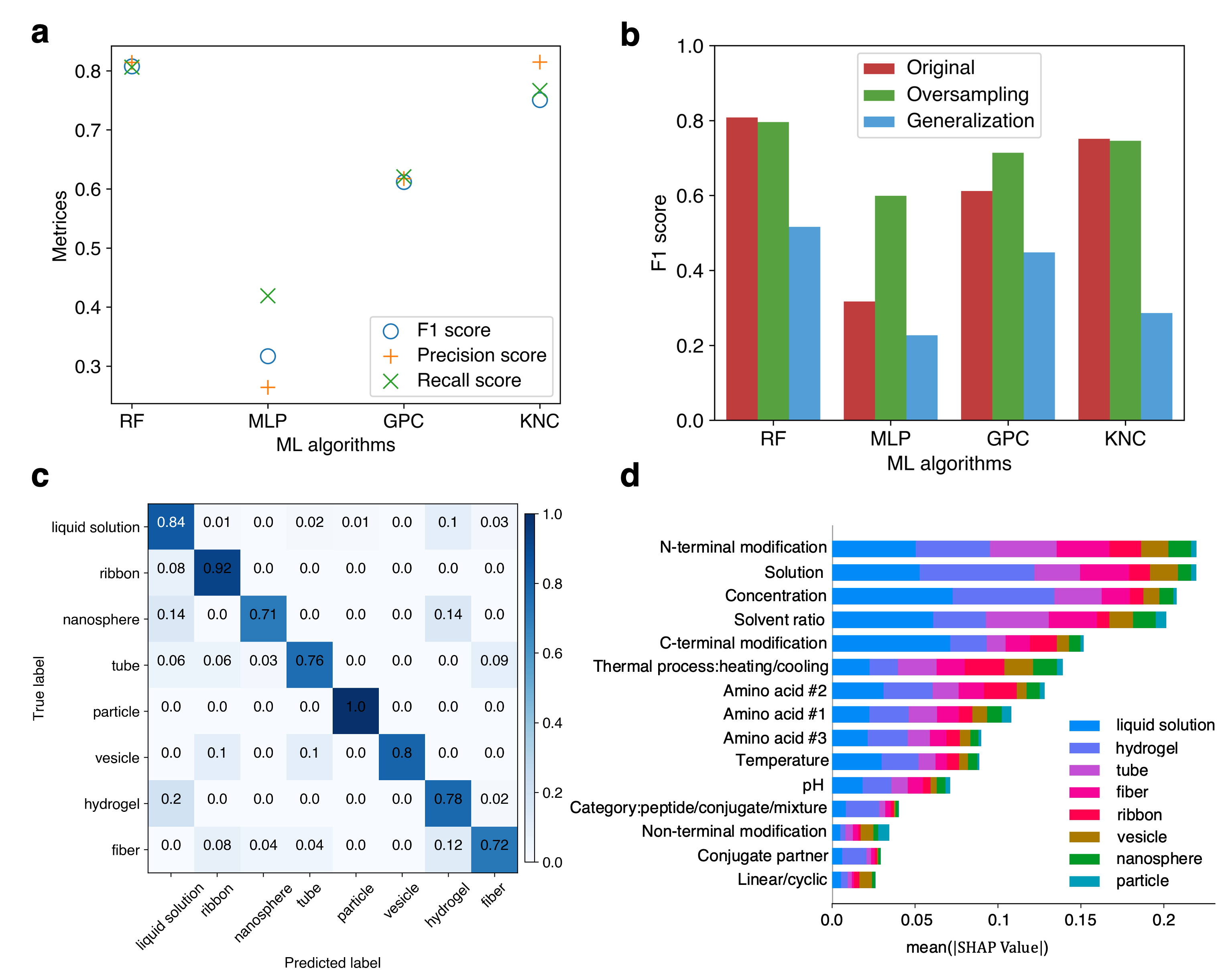}
\caption{\textbf{ML algorithms for phase prediction}. \textbf{(a)} Performance comparison of 4 classical ML classifiers; RF, MLP, GPC and KNC. $F_1$, precision and recall scores are utilized as metrics for evaluation. \textbf{(b)} Comparison of models' performances on the imbalanced dataset (comparison between "original" and "oversampling" case) and evaluation of generalization capacity (comparison between "original" and "generalization" case). \textbf{(c)} Confusion matrix of RF model for 8 different phases. \textbf{(d)} SHAP plot \cite{NIPS2017_7062} for feature importance analysis. }
\label{fig:phase_prediction}
\end{figure}
\phantomsection
\subsection*{ML algorithms for phase prediction}\label{sec:ml_phase_prediction}
With the curated dataset, we are now able to train ML algorithms for phase prediction with different experimental conditions and peptide sequences. Categorical features are converted to one-hot encodings based on the number of classes. All input features are then concatenated into a 1D vector as input to classic ML algorithms. We compare 4 different classic ML algorithms; random forest (RF), multilayer perceptron (MLP) classifier, Gaussian process classifier (GPC) and K-nearest neighbor classifier (KNC). To optimize the performance of these models, we conduct a grid search of hyperparameters for each of them. More details of hyperparameter selection can be found in Methods section \hyperref[sec:methods_ml]{ML algorithms for phase classification} and the grid search results are displayed in Fig. S1. Considering the dataset's imbalance across eight distinct phases (Fig. \ref{fig:data_stats}c), metrics including precision, recall, and $F_1$ scores are integrated to provide a comprehensive evaluation of 4 different ML algorithms. Among tested ML models, RF exhibits the best performance, achieving the highest score (precision=0.814, recall=0.806 and $F_1$=0.808) across all 3 metrics (Fig. \ref{fig:phase_prediction}a). KNC model ranks second because it exhibits a precision score comparable to that of the RF model and its recall and $F_1$ scores are slightly lower. This indicates that KNC's performance is more easily affected by the imbalance of the dataset. In contrast to the RF and KNC algorithms, the MLP classifer exhibits much worse accuracy, characterized by not only lower overall scores but also a large discrepancy among different metrics.\\
\\
To further evaluate the performance of our ML models, we utilize an oversampling technique known as Synthetic Minority Oversampling Technique (SMOTE) \cite{10.5555/1622407.1622416} to address the imbalance of the dataset. The SMOTE approach generates synthetic examples of the minority class to balance the dataset by interpolating between existing minority instances. When utilizing an oversampled dataset, the $F_1$ scores of the MLP classifier and GPC model see a substantial improvement, whereas the RF and KNC algorithms maintain performances comparable to those observed with the original dataset (Fig. \ref{fig:phase_prediction}b). Among the 4 models evaluated, the RF model consistently demonstrates the highest accuracy when tested with a synthetic balanced dataset. However, our current method of randomly splitting the dataset may introduce bias, as both the training and testing sets could contain data from the same publication. This is because a single paper often contributes multiple data entries to the dataset. To evaluate the generalization ability of our models, we partition the dataset into training and testing sets based on publications, which ensures that the data in the testing set originates from publications completely unseen by the training set. As illustrated in Figure \ref{fig:phase_prediction}b, the $F_1$ scores of models evaluated on the generalized test set are dramatically lower than in the original scenario. However, the RF model still exhibits a relatively high $F_1$ score (greater than 0.5), noteworthy considering that a random phase guess would only achieve a score of 0.125, given the 8 phase categories in total. \\
\\
Based on thorough evaluation of the models, we conclude that the RF model is the optimal choice for phase prediction, given its consistently superior performance across all tests. Fig. \ref{fig:phase_prediction}c shows the confusion matrix predicted by the RF model, which compares the true and predicted classifications, visualizing the counts of correct and incorrect predictions across different categories. The high values observed along the diagonal of the confusion matrix underscore model's effectiveness and consistency in classification across different phases. All other confusion matrices for different ML algorithms tested with different situations are listed and compared in Fig. S2. \\
\\
After identifying an optimal model and assessing its performance, we aim to delve deeper into understanding how various input features—from peptide sequences to external stimuli—impact the final output, the self-assembly phase. To acquire interpretability of our model, we employ the SHapley Additive exPlanations (SHAP) technique \cite{NIPS2017_7062} which is a game theoretic approach that explains the importance of input features to the output for a ML model. Among all input features, "N-terminal modification:, "solution" and "concentration" are the top 3 most influential features for phase classification overall (Fig. \ref{fig:phase_prediction}d). Among the two most common phases ("liquid solution" and "hydrogel"), "concentration" is the most critical feature for the "no-assembly" phase, whereas "solution" stands out as the paramount feature for the "hydrogel" phase. This is expected given that for a new phase to be nucleated, a critical concentration must be reached. The correlations uncovered by our model offer valuable insights that can inform experimental designs. For instance, researchers can prioritize the most influential features (identified in the SHAP analysis) in their experiments when aiming to achieve a specific phase.

\begin{figure}[ht]
\centering
\includegraphics[width=1.0\linewidth]{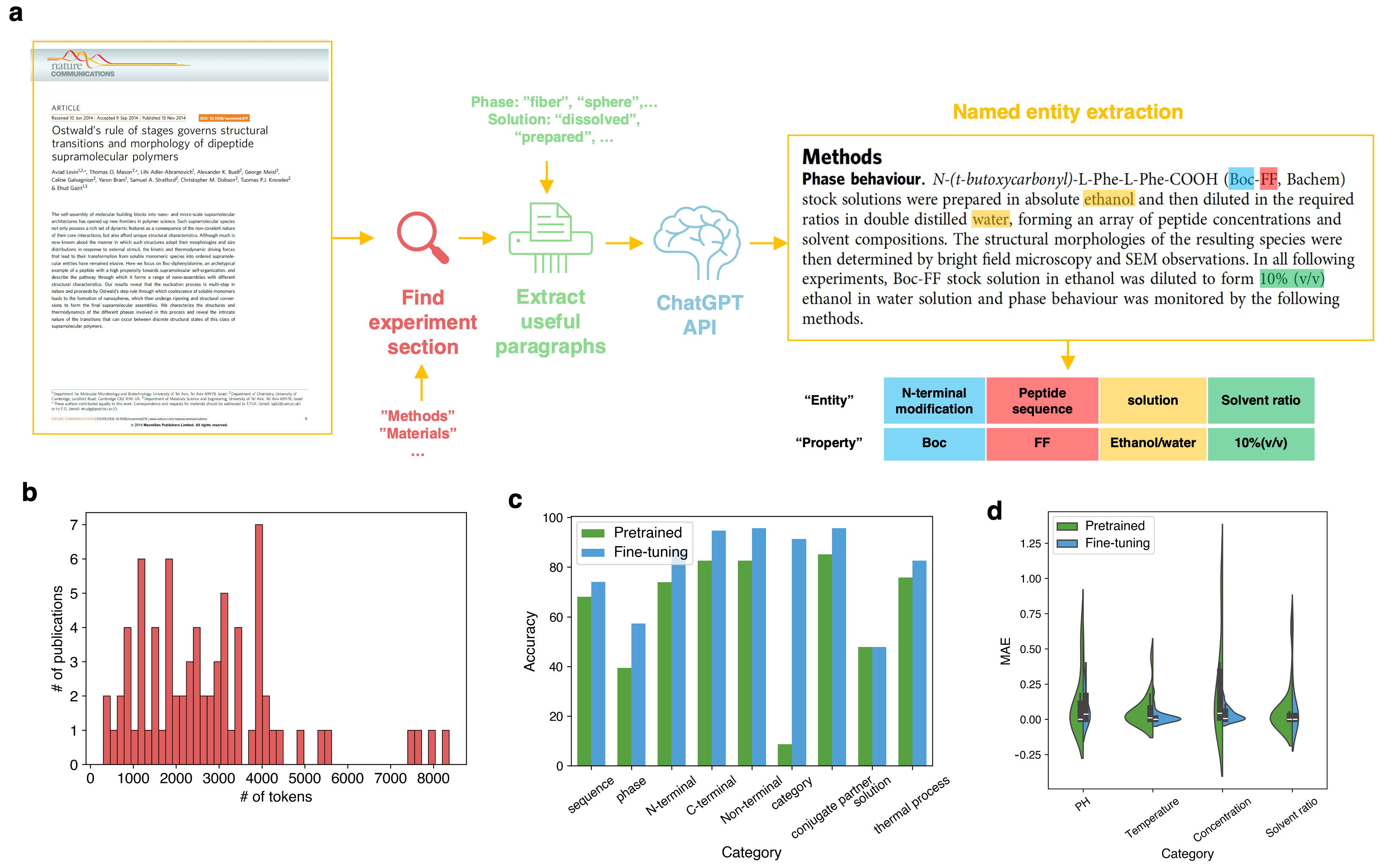}
\caption{\textbf{LLM-assisted literature mining}. \textbf{(a)} Overall workflow of LLM-assisted literature mining: we first extract texts within the experimental section of each publication by searching for the section headings such as "Material(s)". "Method(s)" and "Experimental section(s)". Afterwards, relevant paragraphs are collected based on key words related to target information. For instance, for self-assembly phase, we deliberately search for the name of phases including "fiber", "sphere" and others. With selected paragraphs after preprocessing, we employ both pretrained and fine-tuned (with our manual dataset) to perform "Named Entity Extraction" which extracts 13 target features the from text corpus. \textbf{(b)} Histogram of length of texts after preprocessing. \textbf{(c)} Performance comparison of original pretrained GPT model and fine-tuned model for categorical feature extraction. \textbf{(d)} Performance comparison of original pretrained GPT model and fine-tuned model for numerical feature extraction.}
\label{fig:llm_literature_mining}
\end{figure}

\phantomsection
\subsection*{LLM-assisted literature mining}\label{sec:literature_mining}
While the manual extraction of data from literature by human experts offers high accuracy, it is time-consuming and labor-intensive. To streamline the information extraction process and enhance the efficiency of data collection, we here implement LLMs for automated mining of polypeptide self-assembly literature and further utilize our manually curated database to fine-tune and evaluate the performance of LLMs. The overall workflow of LLM-assisted literature mining is shown in Fig. \ref{fig:llm_literature_mining}a. We first download the PDF files of 75 academic publications from different journal publishers (Fig. S3). The PDF files are then converted to texts using PDFMiner python package \cite{pdfminer2024}. We then locate experimental sections from the whole text document by searching for section headings such as ``Material(s)", ``Method(s)" and "Experimental section/detail(s)". If no section heading is found, all texts are kept for further processing. Afterwards, we gather relevant paragraphs based on the occurrence of key words associated with the information we aim to collect. For example, we collect paragraphs with key words such as ``fibers" and ``hydrogel" to obtain content related to self-assembly phases. Finally, we add the abstract text, which provides a summary of each scientific paper, into the processed main text. We preprocess the texts to reduce the volume of text input, thereby meeting the token (number of words) limits for the training and inference of LLMs, and to augment the efficiency of data mining of literature. \\
\\
The distribution of token numbers for the 75 publications after preprocessing is shown as Fig. \ref{fig:llm_literature_mining}b. We utilize the OpenAI API to call the GPT model \cite{OpenAIGPT3.5Turbo} for information extraction from these processed texts. The task is known as ``Named entity extraction" (NER) which  identifies and classifies key information elements from text into predefined categories (Fig. \ref{fig:llm_literature_mining}a). To enhance the model's performance on NER for peptide literature mining, we fine-tune the GPT model with our peptide database. The peptide database is divided into training (52 papers) and testing (23 papers) sets. Given that most publications yield multiple data entries, we incorporate all data entries from the same publication into a single output, allowing the GPT model to better extract multiple sets of target entities from a single paper. We compare the performances of the original and fine-tuned GPT models by evaluating their prediction accuracy on the testing papers from the manually curated dataset. \\
\\
For 9 categorical features, we assess the performance of LLMs as a True-or-False task by comparing features extracted by the LLMs with those identified by human experts. The fine-tuned LLM dramatically outperformed the pretrained model for all 9 categorical features with an average accuracy over 80\%. In comparison, the original GPT model only achieves an average accuracy of 62.7\%. More specifically, fine-tuning significantly enhances the accuracy of predictions for classification queries. For example, the accuracy for identifying the feature ``category: peptide/conjugate/mixture"—which determines whether the peptide system comprises a single peptide, a conjugate, or a mixture—increases from less than 10\% to over 90\% after fine-tuning. However, fine-tuning does not improve the model's ability to extract information for the feature ``solution." This limitation arises from several factors: (1) "solution" often involves multiple chemicals, complicating the capture of comprehensive information; (2) information about the solution is frequently sparse or absent (appear in figures instead) in the processed texts used as input, making accurate extraction challenging. \\
\\
In terms of the numerical features, we calculate the mean absolute error (MAE) of extracted and ground truth values to compare the accuracy for LLMs with and without fine-tuning. As clearly depicted in Fig. \ref{fig:llm_literature_mining}d, the fine-tuned model not only attains a MAE value but also exhibits a narrower error distribution compared to the pretrained model. This demonstrates that fine-tuning significantly enhances both the accuracy and robustness of the GPT-3.5 turbo model for peptide literature mining. With our fine-tuned LLM, we can significantly enhance the efficiency of dataset curation for learning rules of peptide self-assembly. Furthermore, this workflow can be adapted to other fields of the physical sciences, as experimental conditions are crucial for scientific investigations.

\phantomsection
\section*{Discussion}
In this work, we manually collect data from academic literature to construct a dataset of peptide self-assembly systems, with a particular emphasis on peptide sequences and experimental conditions. Utilizing the database, classical ML algorithms, such as the RF model, are employed to predict the phases of assembled nanostructures based on information regarding both internal chemical composition and external experimental stimuli. The model demonstrates high accuracy in classifying various phases and exhibits moderate generalization capabilities. By further interpreting the classification results, the importance of different experimental conditions to individual phases can be evaluated. These insights offer valuable guidance for researchers in designing experiments aimed at creating specific phases of peptide nanostructures. \\
\\
We further replace the labor-intensive manual data collection process with an LLM assistant. Compared to human experts, the LLMs are capable of processing academic texts and extracting target entities in an automate and efficient manner. We utilized the giant pre-trained GPT model and demonstrated that by fine-tuning it with only a small set of publications, the LLM assistant can achieve impressive accuracy in extracting experimental information, significantly surpassing the performance of the original GPT model designed for general human language tasks. This fine-tuned model will facilitate the streamlining of the data collection process for future publications. The extracted data can subsequently be integrated into our database, further refining the ML algorithm trained for phase classification. Despite the notable improvements in accuracy achieved through fine-tuning our language model, it's important to recognize the existing limitations of our LLM-assisted literature mining approach, which present opportunities for further investigation and refinement in future studies. The major limitations and potential solutions to them are shown in Fig. S4 and Table S4.\\
\\
The proposed workflow and developed models from this work can help us deepen our understanding of the self-assembly of peptide materials. However, it is important to note that the experimental data extracted from literature tends to be biased towards specific peptides (as shown in Fig. \ref{fig:data_stats}d, e) or experimental methodologies (as shown in Fig. \ref{fig:data_stats}a, b, c). Therefore, future efforts could be directed towards incorporating data from domains that have been less explored in previous experiments. Multiple methodologies can be employed to achieve this objective. For example, given the costs of conducting experiments, an active learning-based framework could be adapted to identify a minimal number of experiments required to extend knowledge beyond the current database. High-throughput screening is another potential way of conducting massive experiments in a systematic and automate manner \cite{Marchesan2013}. Compared to experiments, computational tools such as coarse-grained MD simulations are generally much faster but less accurate in phase prediction. Nevertheless, we can harness the strengths of both experimental (high-fidelity but limited in quantity) and computational (low-fidelity but abundant) techniques through the application of multi-fidelity learning \cite{Fare2022}. This approach allows the integration of both experimental and computational data to train ML models with both high accuracy and data efficiency.
Apart from physics-based experimentation and modeling, generative modeling also serves as a useful method for data generation. Model architectures from variational autoencoder, generative adversarial networks to diffusion models are widely applied to augment the data pool of materials and proteins \cite{doi:10.1126/science.aat2663, 10.1063/5.0082338}. These approaches can be easily adapted to produce new experimental data of peptide self-assembly by learning from the existing data we have collected. Further provided with certain conditions like desired phase or constrained experimental parameters, a conditional generative model is a promising approach to assisting experimental design as we discussed in earlier sections. \\
\\
Finally, while this study primarily concentrates on extracting information about the self-assembly phase of peptide materials, the overarching workflow can be applied more broadly. For instance, as a structured functional material, tuning the self-assembly process is essential for accessing diverse applications. Leveraging our fine-tuned LLM for literature mining and the ML classifier for phase prediction, it becomes straightforward to gather property information about peptides from the literature and to construct a model for predicting their functions. Given the similarity between these tasks, there's no necessity to train these models from scratch; transfer learning suffices for precise data extraction and function predictions. Further with the generative models, we can not only design experimental parameters but also the peptide sequences to realize specific functions. These approaches will largely facilitate designing self-assembling peptide materials without conducting costly experiments. 

\section*{Methods}

\phantomsection
\subsection*{Manual processing of peptide database}\label{sec:methods_data}
We first select the data entries from SAPdb database based on the self-assembly phase. Our database includes the top 7 most frequently occurring phases, 8 inclusive of `no assembly'—hydrogel, fiber, tube, sphere, particle, ribbon and vesicle—since other phases have too few data entries, likely making them difficult to be accurately predicted by ML algorithms. Including the non-assembly case, there are a total of 8 phases. We also excluded publications that involve rare methods and techniques that can not be incorporated within the 13 features illustrated in Fig. \ref{fig:data_stats}. These publications include following cases: (1) Experiments of co-assembly of multiple peptides such as Ref. \cite{doi:10.1021/acs.biomac.7b00787}; (2) Purely computational study such as Ref. \cite{C3NR02505E}; (3) Studies on peptides that can not split into a sequence of amino acids such as Ref. \cite{BHARDWAJ2016672}; (4) Experimental control beyond solution environment, PH, temperature such as in-situ ultrasound approach \cite{C5CC02049B}, oxidation \cite{doi:10.1021/ja0040417}, electromagnetic field \cite{Baskar2017} ; (5) Experiments in a solution environment with more than one solvents, Ref. \cite{https://doi.org/10.1002/psc.963}; (6) Experiments without information of solvent/peptide concentration, Ref. \cite{doi:10.1073/pnas.1014796108}; (7) Experiments with rare (less than 5 entries in SAPdb) solvents, additives or chemical modifications such as Ref. \cite{Liebmann2007}. \\
\\
With the screening, we obtain 75 publications in total. Typically, each publication yields multiple data entries, as a single study often investigates various combinations of experimental parameters. Here are the key principles we utilize to extract data entries from each publications: (1) For any parameters that fall within a specific range, we uniformly sample six entries within that range to constitute the data; (2) For a given critical value of phase transition, we select two values (1.1$\times$, 1.2$\times$) above the critical value for one phase and two values (0.8$\times$, 0.9$\times$) below the critical value for the other phase; (3) For peptide concentration, we standardize all values to the unit of mg/ml by calculating the molecular weights of peptides; (4) For pH values, if not specified in the publications, we obtain the value based on the solvent and solute. For instance, for pH of methanol-water mixture in Ref. \cite{Yadav2015}, the pH value is calculated to be 8.3 according to the volume ratio; (5) For temperature, if not specified in the publication, we presume that the experiments are performed under the room temperature (25\textcelsius). By applying these rules to the publications in our database, we derive 1012 data entries. 

\phantomsection
\subsection*{ML algorithms for phase classification}\label{sec:methods_ml}
The input feature vector for phase classification is composed of two parts: the first from categorical features, represented through one-hot encodings, and the second from numerical features. For categorical features with multiple elements, such as "peptide sequence" and "solution", each element is converted into a one-hot encoded vector, and these vectors are then concatenated to form a single 1D feature vector. These categorical and numerical features are combined to create an input feature vector, which has a total length of 139. The output we train the ML algorithms to predict is the self-assembly phase which corresponds to 8 different classes.  \\
\\
To optimize the performance of the 4 classical ML algorithms, grid search of hyperparameters is performed. Grid search is a method used to identify the optimal combination of hyperparameters by systematically exploring a specified search space, which is structured as a discrete grid. For each ML algorithm, we choose 3 independent hyperparameters to form the grid as shown in Table S2. 5-fold cross validation is implemented to compare the models under different combinations of hyperparameters. The whole curated dataset is randomly split into training (60\%), validation (15\%) and testing (25\%) sets. All codes for hyperparameter optimization and ML calculation are written using scikit-learn package \cite{scikit-learn}. The results of grid search along with the optimal hyperparameter combination for 4 ML models are visualized in Fig. S1. \\
\\
In terms of the SMOTE oversampling technique, we implement it using the imbalanced-learn python library \cite{JMLR:v18:16-365}. After oversampling, the size of the training set increases from 607 to 842 with interpolated data in minority classes. To test the generalization capacity of our ML algorithms, we split the 75 publications instead of the whole dataset into training (45 papers), validation (11 papers) and testing sets (19 papers). As a result, the final numbers of data entries for training, validation and testing set are 605, 151, 256 respectively. To better interpret the impact of different features on the self-assembly phase, we implement the SHAP technique using the official SHAP python package \cite{NIPS2017_7062, lundberg2020local2global}. Given that categorical features are represented by one-hot encodings which span multiple dimensions in the input vector, we group all dimensions related to the same feature by summing up their SHAP values. 

\phantomsection
\subsection*{Text processing of literature}\label{sec:methods_preprocessing}
We first download PDF files of 75 publications in our database and then convert the files to text. No images, tables and videos are included after the conversion and we exclude supporting information/supplementary materials given the text volume limit. As we are interested in the effects of experimental conditions on the self-assembly process of peptides, we then extract experimental sections based on the section headings. These headings include:  "Material(s)", "Method(s)", "Material(s) and method", "Method(s) and material(s)", "Experimental section(s)" and "Experimental detail(s)". If none of them are found in the main text, the complete text document is kept for the following processing. To determine the end of the text within the experimental sections, we search for the headings of sections that appear immediately after these experimental parts. These headings include: "Result(s) and discussion(s)", "Result(s) and conclusion(s)", "Result(s)", "Discussion(s)", "Conclusion(s)", "Literature cited", "Acknowledgement(s)" "Reference(s)", "Associated content(s)", "Author information" and "Conflict of Interest(s)". If none of them are found, the text after the experimental section will all be included. \\
\\
With the text corpus of experimental details, paragraphs with target information are then collected. The target information is searched based on a vocabulary of key words which are relevant to the input features and output phase. The list of key words is displayed in Table S3. If any of key words are found within a paragraph, that paragraph is kept in the text for data mining; otherwise, it is removed. After collecting the relevant paragraphs, we eliminate these extremely short paragraphs (with less than 100 characters) and lines (single line with less than 9 characters) which are present mainly due to formatting errors. Upon completing these preprocessing steps, the final token count for 75 papers is presented in Figure \ref{fig:llm_literature_mining}b. Considering the token limit for fine-tuning the GPT model, for processed texts that exceed this limit, we cut the beginning and ending portions to adhere to the token constraint. These texts are generally those without clear headings for different sections. This approach is taken because the beginning of a paper is the introduction section, which generally contains less information about the study itself. Similarly, the ending is often conclusions and discussions, and is less critical compared to the main text.

\phantomsection
\subsection*{LLM for literature mining}\label{sec:methods_llm}
In this work, we employ the "GPT-3.5-turbo-0125" version of GPT model \cite{OpenAIGPT3.5Turbo} for both pretrained and fine-tuned cases considering the accuracy and computational costs. The maximum number of tokens that can be utilized for fine-tuning is 4096. Therefore, texts with more than 4096 tokens are truncated to meet the limit. We pass the name of both categorical and numerical features to our LLMs as entity prompts to perform the NER task. The LLM is fine-tuned with batch size of 1 for only 5 epochs until the loss reaches a plateau. \\
\\
Once the LLM extracts the target information from the text, we proceed to assess its accuracy manually using the following protocol: (1) For every data entry predicted by the LLM, we search for data in our manually curated dataset (considered as the ground truth) that aligns with the predicted peptide sequence and self-assembly phase. (2) In cases where multiple or no direct matches are found, we evaluate the remaining target entities apart from the sequence and phase, using the ground truth data that most closely match the LLM’s predictions for accuracy calculation. (3) For categorical features, the prediction is considered accurate only if prediction and ground truth categories fully align with each other. For features with multiple elements \textit{e.g.} "solution", both solute and solvent need to be correctly identified for the prediction to be considered True. (4) For numerical features, accuracy is determined by comparing the predicted and true values using the MAE. Entries lacking specific numerical values (indicated by "nan" where no values are extracted) are excluded from the MAE calculation. 

\bibliography{sample}

\section*{Acknowledgements and funding}
M.J.B. and Z.Y. acknowledge support from USDA (2021-69012-35978), DOE-SERDP (WP22-S1-3475), ARO (79058LSCSB, W911NF-22-2-0213 and W911NF2120130) as well as the MIT-IBM Watson AI Lab, MIT’s Generative AI Initiative, and Google. Additional support from NIH (U01EB014976 and R01AR077793) ONR (N00014-19-1-2375 and N00014-20-1-2189) is acknowledged. S.K.Y acknowledges funding from the EPSRC Cambridge NanoDTC (EP/S022953/1) and Cambridge Display Technologies Ltd. T.P.J.K. acknowledges funding from the European Research Council under the European Union’s Seventh Horizon 2020 research and innovation program through the ERC grant DiProPhys (agreement ID 101001615), the Biotechnology and Biological Sciences Research Council (BBSRC), the Frances and Augustus Newman Foundation, and the Centre for Misfolding Diseases.

\section*{Conflict of interest declaration}

The authors declare no conflicts of interest.

\section*{Author contributions statement}

M.J.B., T.P.J.K., Z.Y. and S.K.Y. conceived the idea. Z.Y. and S.K.Y. collected data and developed the database. Z.Y. developed and evaluated the ML algorithms and LLMs. M.J.B. and T.P.J.K. supervised the project, analyzed the results, and interpreted them with Z.Y. and S.K.Y.. Z.Y., S.K.Y., T.P.J.K. and M.J.B. wrote and revised the manuscript. 

\section*{Additional information}

\textbf{Supporting information}: The supplementary information of this study can be found at: \textbf{Accession codes}: Dataset and codes created and used in this study can be found at: \href{https://github.com/lamm-mit/PeptideMiner}{https://github.com/lamm-mit/PeptideMiner}; \textbf{Competing interests}: The authors declare that they have no competing interests.

\clearpage


\section*{Supplementary material}

# Supporting Information: Learning the rules of peptide self-assembly through data mining with large language models

**Zhenze Yang**[1,2]**, Sarah K. Yorke**[3]**, Tuomas P. J. Knowles**[3,4,*]**, and Markus J. Buehler**[1,5,6,*]

[1]Laboratory for Atomistic and Molecular Mechanics, Department of Civil and Environmental Engineering, Massachusetts Institute of Technology, 77 Massachusetts Ave. Room 1-165, Cambridge, 02139, MA, USA
[2]Department of Materials Science and Engineering, Massachusetts Institute of Technology, 77 Massachusetts Ave., Cambridge, 02139, MA, USA
[3]Yusuf Hamied Department of Chemistry, University of Cambridge, Lensfield Road, Cambridge, CB2 1EW, UK
[4]Cavendish Laboratory, University of Cambridge, J J Thomson Avenue, Cambridge, CB3 0HE, UK
[5]Center for Computational Engineering, Massachusetts Institute of Technology, 77 Massachusetts Ave., Cambridge, 02139, MA, USA
[6]Center for Materials Science and Engineering, Massachusetts Institute of Technology, 77 Massachusetts Ave., Cambridge, 02139, MA, USA
*mbuehler@mit.edu, tpjk2@cam.ac.uk

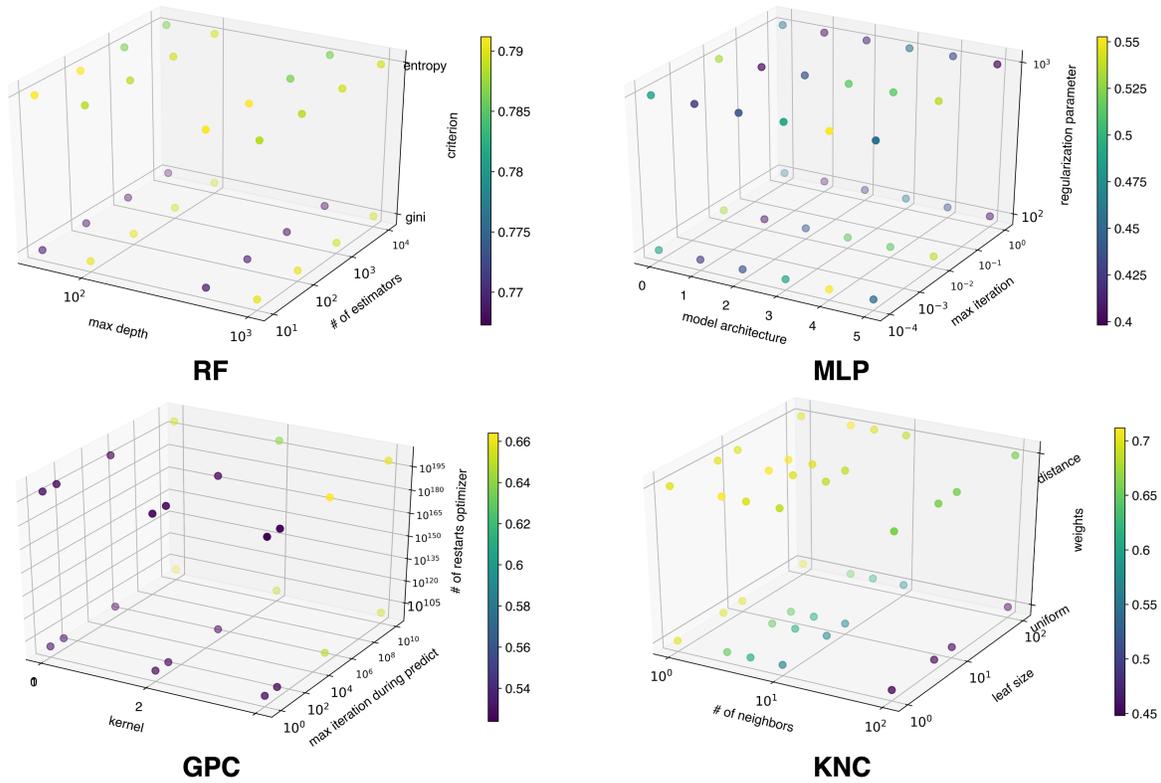

**Figure S1. Hyperparameter grid search for ML algorithms**. The hyperparameter combination for optimal RF model is: 'max depth': 50, '# of estimators': 100, 'criteion': 'gini'; for optimal MLP classifier is: 'regularization parameter': 0.0001, 'model architecture': (128, 256, 128, 16), 'max iteration': 100; for optimal GPC model is: 'kernel': 2RBF(1), 'max iteration during predict': 100, '# of restarts optimizer': 1; for optimal KNC model is: 'leaf size': 1, '# of neighbors': 3, 'weights': 'distance'.

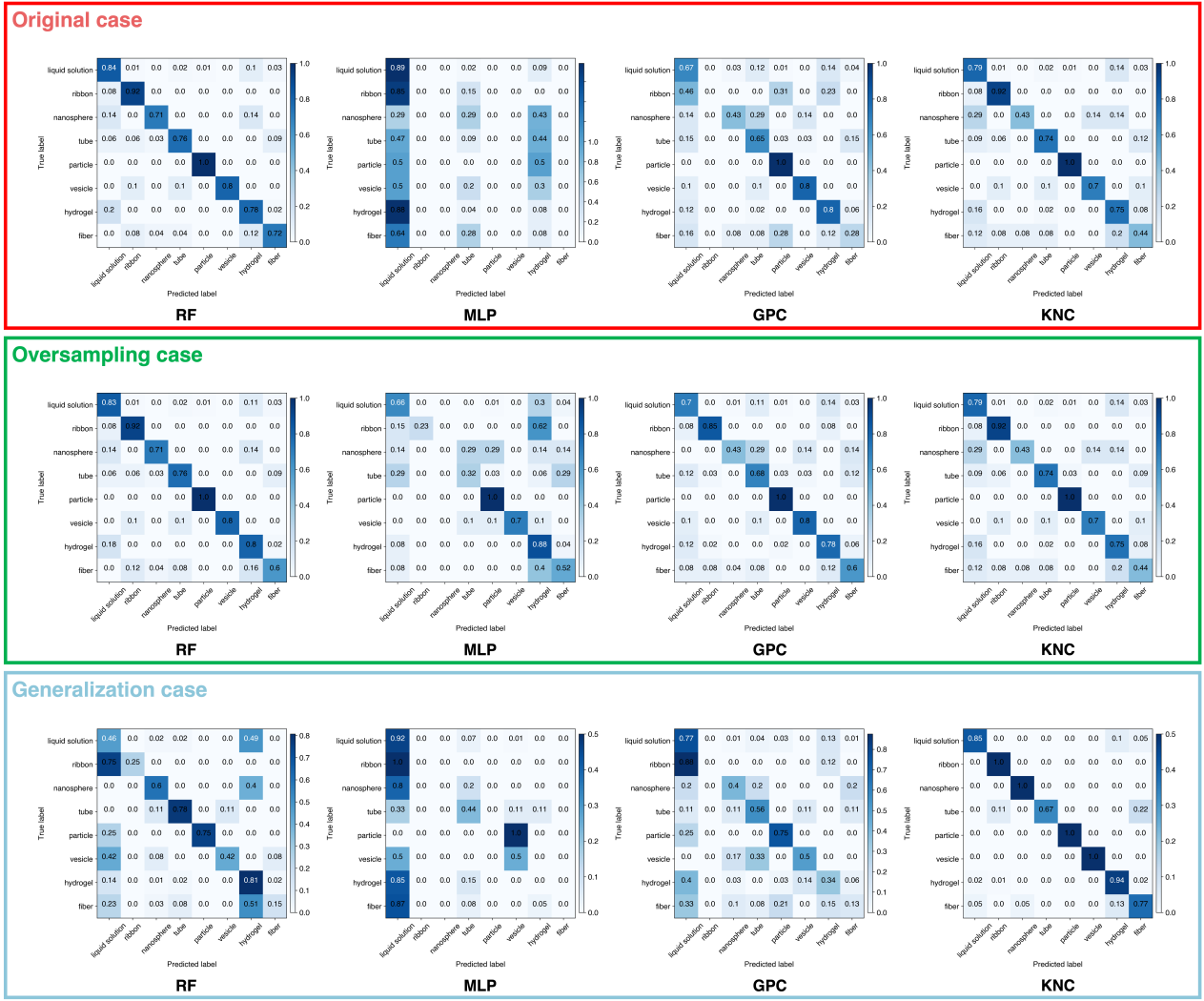

**Figure S2. Confusion matrices**. The confusion matrices, which detail the performance outcomes depicted in Fig. 3b, are arranged in a systematic manner: from top to bottom, they are the original case, the oversampling case, and the generalization case. Horizontally, from left to right, the matrices correspond to the RF, MLP, GPC, and KNC models.

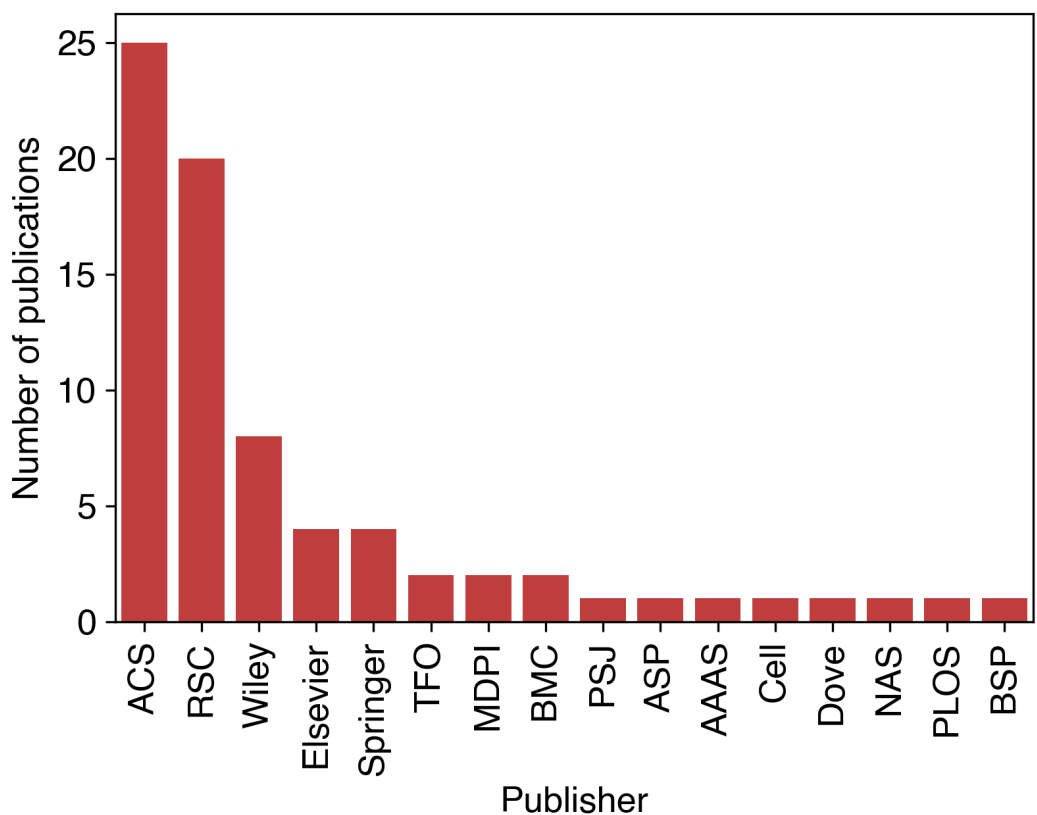

**Figure S3. Publisher information**. We here list the number of publications[1–75] from different journal publishers. The publishers' abbreviations stand for: "ACS" - "American Chemical Society"; "RSC" - "Royal Society of Chemistry"; "TFO" - Taylor & Francis Online; "MDPI" - "Multidisciplinary Digital Publishing Institute"; "BMC" - "BioMed Central"; "PSJ" - "Pharmaceutical Society of Japan"; "ASP" - "American Scientific Publisher"; "AAAS" - "American Association for the Advancement of Science"; "NAS" - "National Academy of Sciences"; "BSP" - "Bentham Science Publishers". Based on the histogram, it is noticeable that the publications are mostly from ACS and RSC publishers.

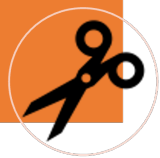
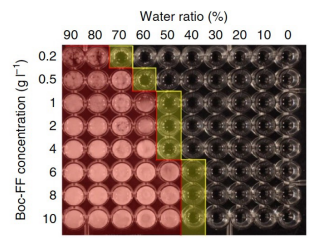
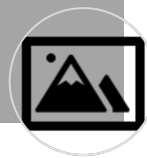
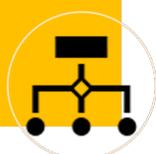
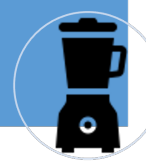

**Figure S4. Limitations of this work**. There are 4 main limitations of current work (from left to right): (1) Due to truncated texts, there can be missing information; (2) No tables or images are included which are frequently used to show self-assembly results; (3) Same group of information can be sparse across the whole text which increases the difficulty of text mining; (4) Information is often mixed. For instance, multiple experiments are utilized to investigate the peptide system which results in multiple sets of experimental conditions. The figure in the limitation related to "No images and tables" is reproduced with permission from Ref.[23]. The table in the same section is created based on the data from Ref.[71]

| Data index | PMID | Peptide sequence | N-terminal modification | C-terminal modification | Non-terminal modification | Category: peptide/ conjugate/mixture | Conjugate partner | Thermal process: heating/cooling |
|---|---|---|---|---|---|---|---|---|
| 1 | 23871602 | FFK | Acetylation | Amidation | None | Peptide | None | No |
| 2 | 23871602 | FYK | Acetylation | Amidation | None | Peptide | None | No |
| 3 | 23871602 | YFK | Acetylation | Amidation | None | Peptide | None | No |
| 4 | 18070345 | FF | Fmoc | None | None | Peptide | None | No |
| 5 | 18070345 | FF | Fmoc | None | None | Peptide | None | No |
| 6 | 18070345 | FF | Fmoc | None | None | Peptide | None | No |
| 7 | 18070345 | FF | Fmoc | None | None | Peptide | None | No |
| Data index | PMID | Linear/Cyclic | Solution | Solvent ratio | Concentration | PH | Temperature | Phase |
| 1 | 23871602 | Linear | HFP/H2O | 0.05 | 1 | 5.5 | 25 | tube |
| 2 | 23871602 | Linear | HFP/H2O | 0.05 | 5 | 5.5 | 25 | fiber |
| 3 | 23871602 | Linear | HFP/H2O | 0.05 | 5 | 5.5 | 25 | fiber |
| 4 | 18070345 | Linear | DMSO/PBS | 0.05 | 5.025 | 7 | 37 | none |
| 5 | 18070345 | Linear | DMSO/PBS | 0.1 | 10.05 | 7 | 37 | none |
| 6 | 18070345 | Linear | DMSO/PBS | 0.2 | 5.025 | 7 | 37 | none |
| 7 | 18070345 | Linear | DMSO/PBS | 0.4 | 10.05 | 7 | 37 | none |

**Table S1. Data examples.** 7 example data entries from two academic publications (PubMed IDs: 23871602 and 18070345).

| Model | RF | | MLP | | GPC | | KNC | |
|---|---|---|---|---|---|---|---|---|
| | Hyperparameter | Values | Hyperparameter | Values | Hyperparameter | Values | Hyperparameter | Values |
| #1 | max depth | 50, 100, 500, 1000 | model architecture | (64, 64, 16), (128, 128, 16), (128, 128, 32), (128, 128, 128, 16), (128, 256, 128, 16), (128, 256, 512, 256, 128, 16) | kernel | RBF(1.0), RBF(10.0), 2RBF(1.0) | # of neighbors | 1, 3, 5, 10, 100 |
| #2 | # of estimators | 10, 100, 1000, 10000 | max iteration | 100, 1000 | max iteration during predict | 100, 200 | leaf size | 1, 5, 10, 100 |
| #3 | criterion | "gini", "entropy" | regularization parameters | 0.0001, 0.01, 1 | # of restarts optimizer | 0, 1, 5, 10 | weights | "uniform", "distance" |

**Table S2. Hyperparameter selection.** Hyperparameters for the RF model include: (1) max depth: the maximum depth of the tree; (2) # of estimators: the number of trees in the forest; (3) criterion: the function to measure the quality of a split. For the MLP classifier: (1) model architecture: the sizes of hidden layers as an array; (2) max iteration: maximum number of training iterations; (3) regularization parameters: strength of the L2 regularization term. For the GPC model: (1) kernel: the kernel specifying the covariance function of the Gaussian process; (2) max iteration during predict: the maximum number of iterations in Newton's method for approximating the posterior during predict; (3) # of restarts optimizer: the number of restarts of the optimizer for finding the kernel's parameters which maximize the likelihood. For KNC model: (1) # of neighbors: number of neighbors used for queries; (2) leaf size: leaf size passed to BallTree or KDTree; (3) weights: weight function category.

| Relevant entity | Key words |
|---|---|
| Phase | "gel", "fibr","lamellar", "flower", "fiber", "filament", "spher", "microcrystal", "tube", "rod", "plate","flake", "particle", "vesic", "ribbon", "bead", "sheet", "tape", "belt", "film" |
| PH, Concentration | dissolve", "dilute", "elute", "wash", "dispers","soluble", "mixed", "added", "pH", "prepare", "concentration","mM","μM", "\u00B5M", "μL", "\u00B5L", "µg","\u00B5g", "g/ml", "gyml", "mg/ml", "mgyml", "g/L", "gyL", "kg/L", "kgyL", "w/w", "w/v", "v/v" |
| Solution | "acetic acid", "acetone", "acetonitrile", "acetyl acetone", "2-aminoethanol", "aniline", "anisole", "benzene", "benzonitrile", "benzyl alcohol", "1-butanol", "2-butanol", "i-butanol", "2-butanone", "t-butyl alcohol", "carbon disulfide", "carbon tetrachloride", "chlorobenzene", "chloroform", "cyclohexane", "cyclohexanol", "cyclohexanone", "di-n-butylphthalate", "1,1-dichloroethane", "1,2-dichloroethane", "diethylamine", "diethylene glycol", "diglyme", "dimethoxyethane", "glyme", "N,N-dimethylaniline", "dimethylformamide", "DMF", "dimethylphthalate", "dimethylsulfoxide", "DMSO", "dioxane", "ethanol", "ether", "ethyl acetate", "ethyl acetoacetate", "ethyl benzoate", "ethlyene glycol", "glycerin", "heptane", "1-heptanol", "hexane", "1-hexanol", "methanol", "methyl acetate", "methyl t-butyl ether", "MTBE", "methylene chloride", "1-octanol", "pentane", "1-pentanol", "2-pentanol", "3-pentanol", "2-pentanone", "3-pentanone", "1-propanol", "2-propanol", "pyridine", "tetrahydrofuran", "THF", "toluene", "water", "H2O", "D2O", "o-xylene", "p-xylene", "m-xylene", "heavy water", "water, heavy", "phosphate-buffered saline", "PBS", "sodium phosphate", "hexafluoro-2-propanol", "HFP", "HFIP", "phosphate buffer", "NaOH", "NaCl", "HCl", "glucono delta-lactone", "GdL", "dichloromethane", "1,3-butanediol", "dimethyl carbonate", "pyrimidine", "pyridine", "mesitylene", "octane", "decane", "isopropanol", "glycine". |

**Table S3. Key words for paragraph collection**. For each paragraph in the text, it is kept if any of the specified keywords are found within it; otherwise, the paragraph is removed.

| Limitations | Solutions |
|---|---|
| Given the limit of token numbers, we trunk the full texts of academic publications which can eliminate important information from the text. | We believe that as training and fine-tuning LLMs become widely accessible to researchers, this issue will be effectively mitigated. |
| We only keep the text of publications for data mining. However, figures, tables and videos often contain essential information about self-assembly results. | Multimodal LLMs are capable of simultaneously processing text, images, and videos, offering a comprehensive and integrated approach to literature mining. |
| The experimental information is often distribute sparsely throughout the full text. For instance, the same group of information, such as observations from electron microscopic examination (see Fig. S4), may be dispersed across various sections of a paper, including Methods, Results, and figure captions. | A potential solution is to group these relevant information by matching key words using another LLMs during the initial stage of preprocessing. Consequently, the LLMs employed for text mining can more effectively learn from text that is originally scattered throughout the document. |
| There are often frequent occurrence of multiple experiments conducted on the same peptide system to explore its morphology, composition, and properties. This result in a mixture of target information we aim to extract, involving multiple sets of experimental conditions which confuses the LLMs we utilize. | We propose the inclusion of the experimental method as an additional entity for information extraction. This approach facilitates the clearer separation of intertwined information. |

**Table S4. Limitations and solutions.** Limitations of current literature mining approach and potential solutions for the future works.


\end{document}